\documentclass[aps,amsmath,preprint,showpacs]{revtex4}
\usepackage{graphicx}
\usepackage{dcolumn}
\usepackage{color}
\usepackage{epstopdf}
\long\def\Omit#1{}

\renewcommand*{\eqref}[1]{Eq.~(\ref{eq:#1})}

\newcommand*{\figlab}[1]{\label{fig:#1}}

\newcommand{\beq}{\begin{eqnarray}}
\newcommand{\eeq}{\end{eqnarray}}

\begin{document}

\title{Charged $K^{*}$ Photoproduction in a Regge model}
\author{Sho Ozaki$^{1}$, Hideko Nagahiro$^{2}$ and Atsushi Hosaka$^{1}$}
\affiliation{1.Research Center for Nuclear Physics (RCNP),
Osaka University, Ibaraki, Osaka, 567-0047, Japan}
\affiliation{2.Department of Physics, Nara Women's University,
Nara 630-8506, Japan}

\date{\today}

\begin{abstract}

We investigate $\gamma p \to K^{*+} \Lambda$ reaction within a Regge approach.
For the gauge invariance of the scattering amplitude,  we reggeize the $s$-channel and contact term amplitudes
as well as the $t$-channel amplitude.
We obtain the decreasing behavior of the total cross section as the CLAS's preliminary data show.
We also calculate spin density matrices, and find clear differences between our Regge model and the previous Feynman (isobar) model.

\end{abstract}
\maketitle

\newpage

Recently strangeness photoproduciton has been of interest in hadron physics.
Several photon facilities such as CLAS, LEPS, SAPHIR and CBELSA/TAPS provide accurate data of strangeness photoproduction.
In these data there are very interesting phenomena, for instance, pentaquark production \cite{Nakano:2003qx, Nakano:2008ee}, peak structures near the threshold  in $\phi$ photoproduction \cite{Mibe:2005er} and also in $\Lambda$ resonance photoproduction \cite{Niiyama:2008rt, Kohri:2009xe}.
To understand those phenomena, we have to understand a fundamental mechanism of open strangeness photoproduction.

From a theoretical point of view, Regge phenomenology is very successful in strangeness photoproduction above $2$ GeV. The model reproduces the energy- and $t$-dependence of cross sections and spin observables of several strangeness photoproduction in the forward angle region. Especially what we pay attention to is the energy-dependence of cross sections predicted by this model.
The Regge phenomenology successfully reproduces the decreasing behavior of cross sections of the kaon photoproduction in terms of kaon and $K^{*}$ trajectories \cite{Mart:2004au, Corthals:2005ce, Corthals:2006nz, Guidal:1997hy}.
On the other hand, in $\phi$ photoproduction, the experimental data show a monotonically increasing behavior of the cross section. This behavior is explained by the Pomeron exchange \cite{Titov:2007fc, Ozaki:2009mj}.

\begin{figure*}
\begin{minipage}{0.65\hsize}
\begin{center}
\includegraphics[width=0.7 \textwidth]{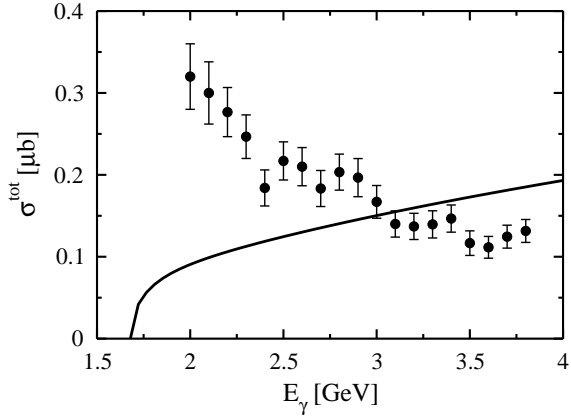}
\end{center}
\end{minipage}
\caption{
Total cross section of the $\gamma p \to K^{*+} \Lambda$ reaction as function of the photon energy $E_{\gamma}$ in the laboratory frame. The solid line is the result of the Feynman model (see text for details), and the data are taken from \cite{Guo:2006kt, Oh:2006hm}.
}
\label{total}
\end{figure*}

Originally, however, the Regge model is applied in high energy hadron reactions.
The relevant question is then at what energy and beyond we can apply the Regge model.
From the successes of applying the model in strangeness photoproduciton, one can expect that it is valid in the energy region where open strangeness is produced.
To test this, we study the charged $K^{*}$ photoproduction by using the Regge approach.
The feature of the charged $K^{*}$ photoproduction is that $K^{*}$ (vector meson) exchange is allowed, whereas it is forbidden in the neutral $K^{*}$ photoproduction.
Since the exchange of a higher spin particle including a vector meson leads to an amplitude increasing with the energy, the Feynman (isobar) model violates the Froissart bound (unitarity) in the high energy region.
Figure~\ref{total} shows CLAS's preliminary data of $\gamma p \to K^{* +}\Lambda$ and the result of the Feynman model.
Since $K^{*}$ (vector meson) exchange amplitude makes monotonically increasing behavior, it is difficult to reproduce the decreasing behavior of the data if one applies the Feynman model naively.
The contradiction between the data and the result of the Feynman model may indicate that we need to employ the Regge approach in this energy region.                  
Instead of one meson exchange, the Regge model takes into account the exchange of entire family of hadrons with the same quantum number expect spin. The resulting amplitude satisfies the Froissart bound.
Due to the slope parameters of the kaon, $\kappa$ and $K^{*}$ trajectories, one can expect that the Regge model naturally reproduces the decreasing behavior.
In this paper we apply the Regge model in the charged $K^{*}$ photoproduction and compare the results of our Regge model with those of the previous Feynman model.

In the previous studies, Oh $et \ al$. investigated $K^{*}$ photoproduction which was based on the Feynman model \cite{Oh:2006hm, Oh:2006in}.
Oh's model nicely reproduced the experimental data.
The exact form of the amplitudes and coupling constants in the model were given in Refs.\cite{Oh:2006hm, Oh:2006in}.
In the model the hadronic form factor are used at each vertex to take into account size effects of hadrons phenomenologically:
\beq
F(t) = \frac{\Lambda_{t}^{2}-m^{2}}{\Lambda_{t}^{2}-t}, \
F(x) = \frac{\Lambda_{x}^{4}}{(x-M^{2})^{2}+\Lambda_{x}^{4}}, \
x  =  s,u,
\eeq
where $m$ and $M$ are masses of an exchanged meson and baryon.
The sum of the electric term of the $s$-channel, contact term and $K^{*}$ exchange term form a set of the gauge invariant terms, and therefore, the common form factor $F_{c} = 1-(1-F(s))(1-F_{K^{*}}(t))$ was employed for them \cite{Davidson:2001rk}.
Figure 2-(a) shows contributions from each channel in the Feynman model with a cut-off $\Lambda_{t} = 1.15$ GeV for all $t$-channels and a cut-off $\Lambda_{x} = 0.9$ GeV for $s$- and $u$-channels.
As shown in this figure, the origin of the monotonically increasing behavior is the $K^{*}$ exchange and also the contact term contribution.
Therefore if we use the same value of the cut-off parameter $\Lambda_{t}$ for all the $t$-channel contributions, we obtain the energy increasing behavior as shown in Fig.\ref{total}.
To explain the decreasing behavior in Oh's model, $K^{*}$ exchange and the contact term contributions were much suppressed by tuning the cut-off parameter of the form factor (Fig.2-(b)).
Then a dominant contribution became the kaon exchange.
This is a unique feature of this model.
If one would like to reproduce charged $K^{*}$ photoproduction in terms of the Feynman model, one should suppress contributions of a vector meson exchange and a contact term, and enhance a scalar or psudoscalar meson exchange terms.

\begin{figure*}
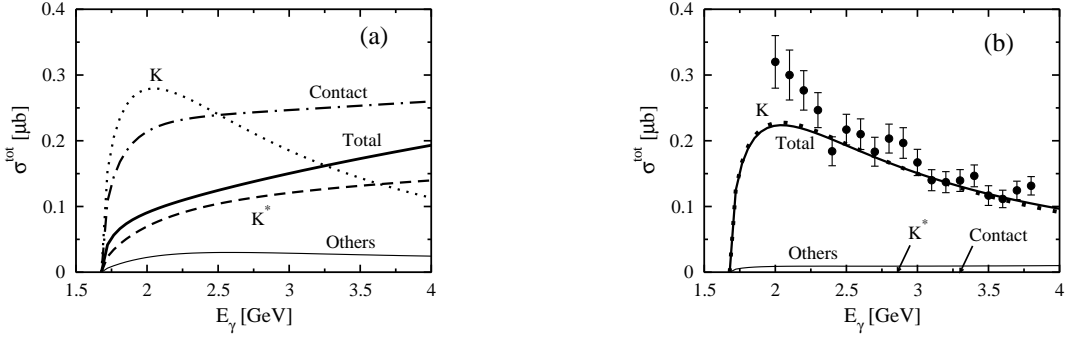

\begin{tabular}{cc}
\begin{minipage}{0.5\hsize}
\begin{center}
\includegraphics[width=0.7 \textwidth]{each_contr_naive_feyn.eps}
\end{center}
\end{minipage}
\begin{minipage}{0.5\hsize}
\begin{center}
\includegraphics[width=0.7 \textwidth]{each_contr_Y-Oh_feyn.eps}
\end{center}
\end{minipage}
\end{tabular}
\caption{
(a) Separate contributions of each cannel to the total cross section in the Feynman model. 
The solid line is total cross section, the dashed line is $K^{*}$ exchange, the dotted line is kaon exchange, the dot-dash line is the contact term and the thin solid line is other channel contributions including $\kappa$ exchange.
(b) The previous results of the Feynman model where the cut-off parameters $\Lambda_{K, \kappa} = 1.1$ GeV and $\Lambda_{K^{*}} = 0.9$ GeV are used in Ref.~\cite{Oh:2006hm}. Contributions  of various terms are separately shown.
Conventions are as in (a).}
\figlab{total}
\end{figure*}

Now we introduce the Regge model for the charged $K^{*}$ photoproduction.
The reggeization is done by replacing the $t$-channel propagator in the Feynman amplitudes by the Regge propagator as
\cite{Corthals:2005ce, Corthals:2006nz, Guidal:1997hy}
\beq
\frac{1}{t-m_{K}^{2}}
&\to& \mathcal{P}_{regge}^{K} = \left(\frac{s}{s_{0}}\right)^{\alpha_{K}(t)}\frac{1}{{\rm{sin}}(\pi \alpha_{K}(t))} \frac{\pi \alpha_{K}^{'}}{\Gamma(1+\alpha_{K}(t))}, \\
\frac{1}{t-m_{\kappa}^{2}}
&\to& \mathcal{P}_{regge}^{\kappa} = \left(\frac{s}{s_{0}}\right)^{\alpha_{\kappa}(t)}\frac{1}{{\rm{sin}}(\pi \alpha_{\kappa}(t))}\frac{\pi \alpha_{\kappa}^{'}}{\Gamma(1+\alpha_{\kappa}(t))}, \\
\frac{1}{t-m_{K^{*}}^{2}}
&\to& \mathcal{P}_{regge}^{K^{*}} = \left(\frac{s}{s_{0}}\right)^{\alpha_{K^{*}}(t)-1}\frac{1}{{\rm{sin}}(\pi \alpha_{K^{*}}(t))}\frac{\pi \alpha_{K^{*}}^{'}}{\Gamma(\alpha_{K^{*}}(t))}.
\eeq
Formally one can reggeize the $t$-channel amplitude by multiplying the $t$-channel Feynman amplitude by
$(t-m^{2}_{K^{*}})\mathcal{P}_{regge}$.
Here we assume degenerate trajectories because there are no dip structures (wrong-signature zeroes) in kaon photoproduction \cite{Corthals:2005ce, Corthals:2006nz} and also in the reaction $\gamma p \to K^{* 0}\Sigma^{+}$ \cite{Nanova:2008kr, Hleiqawi:2007ad}.
We choose constant phases for all trajectories and set $s_{0} = 1$ GeV. Phase dependence will be discussed in Ref.\cite{sho-p}. The following discussions, however, are not affected by this phase factor. Meson trajectories are given by
\beq
\alpha_{K}(t)
&=& 0.70 \ {\rm{GeV}}^{-2}(t-m_{K}^{2}), \\
\alpha_{\kappa}(t)
&=& \alpha_{\kappa}^{'}(t-m_{\kappa}^{2}), \\
\alpha_{K^{*}}(t)
&=& 1 + 0.85 \ {\rm{GeV}}^{-2}(t-m_{K^{*}}^{2}),
\eeq
where $\alpha_{\kappa}^{'}$ is fixed by the $\kappa$ trajectory.
In PDG \cite{Amsler:2008zzb}, $K^{*}_{0}(800)$(or $\kappa$), $K^{*}_{1}(1410)$ and $K^{*}_{2}(1980)$ can be identified as members of the $\kappa$-family.
Other mesons are not found yet.
We can fit $\kappa$ trajectory with $\alpha_{\kappa}^{'} = 0.70 \ {\rm{GeV}}^{-2}$.
To maintain the gauge invariance, we need to reggeize the set of the gauge invariant terms of the electric $s$-channel, contact, and $K^{*}$ exchange terms simultaneously. This means that in accordance with Eq.(4), we perform the following replacement
\beq
({\mathcal{M}}^{elec}_{s} + {\mathcal{M}}_{c})
&\to& ({\mathcal{M}}^{elec}_{s} + {\mathcal{M}}_{c}) \times (t-m_{K^{*}}^{2})\mathcal{P}_{regge}^{K^{*}}
\label{gauge-inv}.
\eeq
This procedure has been shown to be important to reproduce the very forward angle behavior ($\theta \sim 0$) in the charged kaon photoproduction \cite{Mart:2004au, Corthals:2005ce, Guidal:1997hy}.
\begin{figure*}
\begin{minipage}{0.8\hsize}
\begin{center}
\includegraphics[width=1.0 \textwidth]{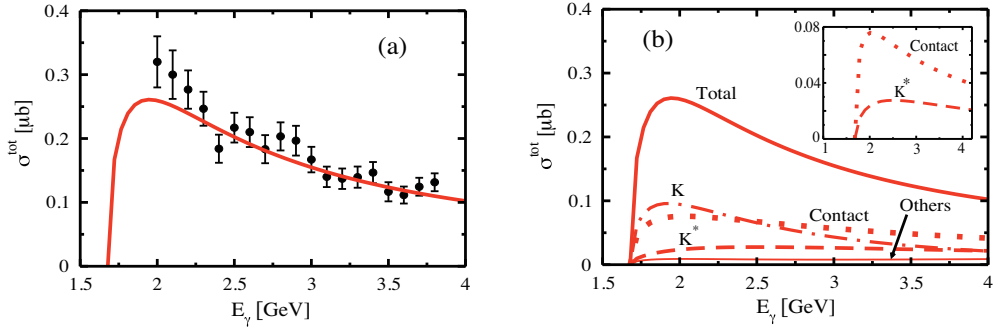}
\end{center}
\end{minipage}
\caption{(a) The total cross section of the reaction $\gamma p \to K^{* +}\Lambda$ in the Regge model.
The cut-off of the $t$-channel is $\Lambda_{t} = 1.55$ GeV.
(b) Various contributions to the total cross section. Conventions are as in Fig.2-(a)}
\figlab{total}
\end{figure*}

Let us now discuss the result of our approach. As shown in Fig.3-(a), we find that the Regge model can reproduce the experimental data very well and describe the decreasing behavior of the total cross section.
Here we use the common cut-off parameter $\Lambda_{t} = 1.55$ GeV for the $t$-channel Regge amplitudes. 
Figure 3-(b) shows each contribution of our model.
In this model $K^{*}$ trajectory and the reggeized contact term contributions are both decreasing as the energy is increased.
Here the gauge invariant prescription Eq.(\ref{gauge-inv}) plays a crucial role in explaining the decreasing behavior.
Although both the Regge and Feynman models reproduce the decreasing behavior with energy, the reaction mechanisms are different between them.

To see the difference more clearly, we calculate spin observables in the reaction $\gamma p \to K^{*+}\Lambda$.
Spin observables are very important to understand the reaction mechanism;
for instance, Ref.~\cite{Ozaki:2007ka, Ozaki:2008zz} pointed out an importance of chiral anomaly in kaon photoproduction by using the photon beam asymmetry.
To focus on the $t$-channel contributions, we study spin density matrices at a forward angle, $\theta = 20$ degrees.
Figure 4-(a) shows $\rho^{0}_{00}$ which represents spin one flip process in the Gottfried-Jackson(GJ) system \cite{Titov:1999eu}.
In this system, a scalar or pseudoscalar exchange makes this matrix element exactly zero, but a vector meson exchange makes a finite contribution to this matrix element.
Since in the Oh's model the dominant contribution is the kaon (psudoscalar meson) exchange, $\rho^{0}_{00}$ is almost zero.
On the other hand, in our Regge model, $\rho^{0}_{00}$ is finite due to the sizable contribution of $K^{*}$ trajectory.
Figure 4-(b) shows $\rho^{1}_{1-1}$ which represents naturalness or unnaturalness of the exchanged particles.
The previous model makes this matrix element almost $-1/2$ due to the kaon exchange (unnatural parity exchange) dominance. 
Since the $K^{*}$ contribution is positive and increasing with the energy, in the Regge model the negative value of $\rho^{1}_{1-1}$ decreases as the energy is increased.
The significantly different behaviors in the two models in the spin density matrices are expected to be observed in future experiments which give us important information on the $K^{*}$ photoproduction mechanism.

In summary, we have studied charged $K^{*}$ photoproduction by using the Regge model, and compared the results of the Regge model and those of the previous one.
Both results reproduce the monotonically decreasing behavior of the total cross section, but the reaction mechanisms are very different between them.
In the Feynman model, the dominant contribution is the kaon (psudoscalar meson) exchange, and $K^{*}$ (vector meson) exchange and the contact term are much suppressed to reproduce the decreasing behavior of the total cross section.
On the other hand, in the Regge model $K^{*}$ trajectory and the reggeized contact term contributions are naturally decreasing and provide sizable contributions. Consequently we have found the decreasing behavior of the total cross section as the experimental data show.
Due to these features of the two models, we have found the clear differences in the spin density matrices.
These differences in the spin density matrices can be used in comparison with experimental data to test the Regge model for open strangeness production.


\


\begin{figure*}
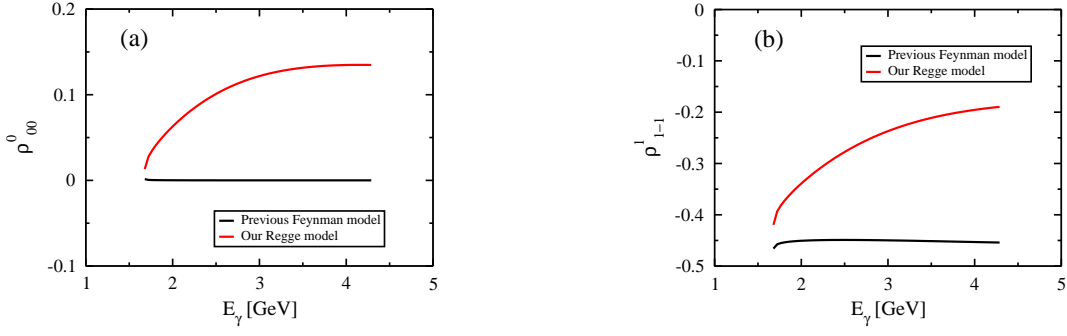

\begin{tabular}{cc}
\begin{minipage}{0.5\hsize}
\begin{center}
\includegraphics[width=0.7 \textwidth]{RZZZ_paper.eps}
\end{center}
\end{minipage}
\begin{minipage}{0.5\hsize}
\begin{center}
\includegraphics[width=0.7 \textwidth]{RPPM_paper.eps}
\end{center}
\end{minipage}
\end{tabular}
\caption{Spin density matrices of $\gamma p \to K^{* +} \Lambda$ reaction,
(a) $\rho^{0}_{00}$ and (b) $\rho^{1}_{1-1}$. In the calculation, the $K^{*}$ angle is fixed at $\theta = 20$ degrees. The red line is our Regge model result, and the black line is the previous Feynman model result.}
\figlab{on_shell}
\end{figure*}



\end{document}